\documentclass{ieeeaccess}
\usepackage{cite}
\usepackage{amsmath,amssymb,amsfonts}
\usepackage{algorithmic}
\usepackage{graphicx}
\usepackage{textcomp}
\usepackage{subcaption}
\usepackage{braket}
\usepackage{url}
\usepackage{booktabs} 
\usepackage{soul,xspace,tabularx}

\newcommand{\eat}[1]{}
\providecommand{\myparab}[1]{\smallskip\noindent\textbf{#1} }
\def\BibTeX{{\rm B\kern-.05em{\sc i\kern-.025em b}\kern-.08em
    T\kern-.1667em\lower.7ex\hbox{E}\kern-.125emX}}
\begin{document}
\history{Received 19 January 2024; revised 16 June 2024; accepted 16 July 2024; date of publication 23 July 2024;
date of current version 22 August 2024.}
\doi{10.1109/TQE.2024.3432390}

\title{Resource Placement for Rate and Fidelity Maximization in Quantum Networks}
\author{\uppercase{Shahrooz Pouryousef}\authorrefmark{1,2},
\uppercase{Hassan Shapourian\authorrefmark{2}, Alireza Shabani\authorrefmark{2}, Ramana Kompella\authorrefmark{2} and Don Towsley}\authorrefmark{1}
\IEEEmembership{Member, IEEE}}
\address[1]{University of Massachusetts Amherst USA
}
\address[2]{Cisco Research, San Jose, CA 
}

\markboth
{Pouryousef \headeretal: Resource Placement for Rate and Fidelity Maximization in Quantum Networks}
{Pouryousef \headeretal: Resource Placement for Rate and Fidelity Maximization in Quantum Networks}

\corresp{Corresponding author: Shahrooz Pouryousef (email: shahrooz@cs.umass.edu). 
}


\begin{abstract}
Existing classical optical network infrastructure cannot be immediately used for quantum network applications due to photon loss. The first step towards enabling quantum networks is the integration of quantum repeaters into optical networks.
However, the expenses and intrinsic noise inherent in quantum hardware underscore the need for an efficient deployment strategy that optimizes the \textcolor{black}{placement} of quantum repeaters and memories. In this paper, we present a comprehensive framework for network planning, aiming to efficiently distribute quantum repeaters across existing infrastructure, with the objective of maximizing quantum network utility within an entanglement distribution network. 
We apply our framework to several cases including a preliminary illustration of a dumbbell network topology and real-world cases of the SURFnet and ESnet. We explore the effect of quantum memory multiplexing within quantum repeaters, as well as the influence of memory coherence time on quantum network utility. We further examine the effects of different fairness assumptions on network planning, uncovering their impacts on real-time network performance.

\end{abstract}




\titlepgskip=-15pt

\maketitle




\section{Introduction}

The advent of the quantum Internet holds immense potential for realizing a wide array of transformative quantum applications, including quantum key distribution (QKD)~\cite{bennett2014quantum,peev2009secoqc,wang2014field,stucki2011long}, quantum computation~\cite{cirac1999distributed,cacciapuoti2019quantum}, quantum sensing~\cite{degen2017quantum}, clock synchronization~\cite{komar2014quantum}, and quantum-enhanced measurements~\cite{giovannetti2004quantum}, among others~\cite{kimble2008quantum}. 
One of the primary challenges to realizing such a large-scale quantum network lies in the transmission of quantum information through optical fiber over long distances, as photon loss increases exponentially with distance. To overcome this limitation, the concept of a quantum entanglement distribution network has been introduced~\cite{briegel1998quantum,munro2015inside,azuma2022quantum}.
The basic idea behind a quantum network is to strategically position a series of repeater stations along the transmission path~\cite{rabbie2022designing,da2023requirements}. By leveraging the concept of entanglement swapping, long-range entangled qubits (in the form of Einstein-Podolsky-Rosen (EPR) pairs) between a pair of end users can be established. This process involves performing Bell state measurements at each intermediate node to effectively combine elementary link entanglements between adjacent repeaters. Once entanglement is established, quantum information can be transmitted through quantum teleportation. Therefore, 
the successful execution of quantum Internet applications demands the development of novel protocols and the integration of quantum hardware, all aimed at establishing and maintaining reliable and high-fidelity entanglement across long distances in a quantum network~\cite{wehner2018quantum,lloyd2004infrastructure,dahlberg2019link,cirac1997quantum,kuhn2002deterministic,liao2018satellite}.






How do we ensure optimal performance of quantum networks in reliably delivering entanglement to the end users? Addressing this question requires a systematic approach, starting with quantum network planning. Similar to its classical counterpart, efficient resource management is crucial in quantum networks. In particular, quantum resources such as quantum repeaters and links must be carefully placed and optimized to meet the specific requirements of user pairs in real-time scenarios. To achieve effective quantum network planning, several key questions need to be addressed. First, determining the optimal number of quantum repeaters and their placement is essential to maximizing the success probability of end-to-end entanglement while maintaining fidelity. 
Additionally, allocating quantum memories at repeaters to users is crucial in achieving network fairness and ensuring efficient utilization of available resources.
Furthermore, the coherence times of quantum memories at both the end-user nodes and repeaters should be accounted for in network planning, as they impose upper bounds on the time frame available for classical communications.


In this paper, we formulate quantum network planning as an optimization problem.
In short, the objective it to maximize quantum network utility with repeater locations as decision variables. Our utility function includes the rate and fidelity of the generated entanglement between end-users. The concept of network utility for classical networks made its debut in the seminal research conducted by Kelly \cite{kelly1997charging,kelly1998rate}. There is a huge amount of research on network utility maximization in classical networks. Analogous to classical network utility, the idea of quantum network utility maximization has been proposed in works such as ~\cite{vardoyan2022quantum,lee2022quantum}. Given a quantum network, Vardoyan~\emph{et al.}~\cite{vardoyan2022quantum} solve an optimization problem for finding the rates and link fidelities in order to maximize the utility function of a set of user pairs. However, in this paper, we start with planning the network for utility maximization.


We study how the following network parameters affect the optimal solution to our network planning optimization problem: number of end-user pairs, distance between network nodes (which can potentially be used as repeaters), repeater capacity (i.e., maximum number of quantum memories per repeater), and quantum memory coherence time. We use a quantum memory multiplexing approach~\cite{azuma2015all,shi2020concurrent} to achieve higher end-to-end entanglement rates and treat memory allocation to different end-user pairs as part of our optimization problem. 
We find that the impact of multi-user demands on the end-to-end entanglement rate becomes more significant as the distance between nodes is increased, while more end users may not necessarily imply the need for more repeaters. We observe that the requirement imposed on coherence time is much less restrictive for repeater memories than it is for end-node memories. 
Finally, we examine the planned network (i.e., the output of our optimization problem) at run-time
given random network traffic and show that its average performance is comparable to an unachievable upper bound. \textcolor{black}{To enable the community to explore our ideas and to facilitate the reproducibility of our results, our code is available online.} \footnote{\url{https://github.com/pooryousefshahrooz/q_net_planning}}


The rest of the paper is organized as follows: In Sec.~\ref{sec:network model}, we introduce our network model and entanglement distribution protocol, and how to characterize the quantum network performance and utility in terms of rate and signal quality. We further explain what is the output of our network planning framework. In Sec.~\ref{sec:opt}, we present our network planning framework as an optimization problem and elucidate two ways of formulating the problem. We discuss why the optimization problem is nonlinear by definition and how we make it linear at the cost of neglecting some effects or introducing extra overhead.
Sec.~\ref{experiment} is devoted to several experiments where we apply our framework to various network topologies. Finally, we conclude in Sec.~\ref{sec:concl} with some closing remarks and future directions. The derivation of the end-to-end entanglement generation rate in the presence of memory multiplexing and some additional optimization results are provided in three appendices.






\begin{figure}
\centering
\includegraphics[scale=0.35]{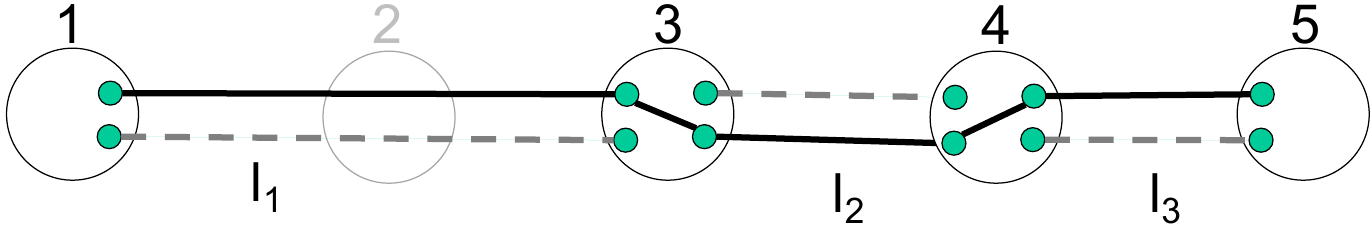}
\caption{An example of quantum network planning for a linear chain with $3$ potential locations for repeaters and a maximum of two memories per path. An instance of end-to-end entanglement generation is shown, where solid (dashed) lines represent successful (failed) attempts on links. The line connecting two memories inside a repeater indicates a successful Bell state measurement. The gray node $2$ shows that no repeater is placed at that location.}
\label{fig:Path_W}
\end{figure}

\section{Network Model}
\label{sec:network model}


We consider a quantum network represented by a graph $G = (V, E)$, where $V$ is the set of nodes and $E$ is the set of optical communication links. {There are two types of nodes in the network: A set of nodes corresponding to the end users denoted by $U \subset V$ and a set of nodes denoted by $R \subset V$ which provides potential locations for placing repeaters} ($V=U\cup R$). \textcolor{black}{Each node $u \in R$  has memory capacity $D_u$. We assume each end node is used by only one user pair and two user pairs can not have the same end nodes. We assume when a quantum repeater is placed at a location with a degree of more than two, it acts as a \textit{quantum switch} to cross connect multiple paths (as introduced in Ref.~\cite{vardoyan2021exact} which is different from the definition in Ref.~\cite{caleffi2020quantum}). We shall still call such nodes a quantum repeater in this paper.} There are $|Q|$ user pairs that we want to maximize their utility function with respect to using at most $N_\text{max}\leq |R|$ repeaters. Unused nodes in $R$ then operate as optical switches.  

We refer to the Einstein-Podolsky-Rosen (EPR) pairs or entanglement bits (ebits) generated along such links as link-level entanglement.
An end-to-end entangled state between a pair of users can be established using a process called \textit{entanglement swapping}, that connects link-level ebits via Bell state measurements (BSMs) at the repeaters. 

For example, suppose two nodes $1$ and $3$ in Fig. \ref{fig:Path_W} share an ebit $\ket{\psi_{13}^+}$, and node $3$ shares another pair $\ket{\psi_{34}^+}$ with node $4$. Then, node $3$ can create an ebit $\ket{\psi_{14}^+}$ between $1$ and $4$ by performing a BSM followed by a classical communication exchange. The process can be repeated to create ebits between distant parties $1$ and $5$. Tables \ref{table:notaions},\ref{table:notaions-path-based}, and \ref{table:notaions-link-based} show the notations used in this paper.

\subsection{Entanglement distribution protocol}
\label{sec:protocol}

We consider a sequential entanglement distribution protocol based on the spatial multiplexing of quantum memories. A path between two users is said to have width $W$ when each end-user has $W$ quantum memories, and each repeater node is equipped with $2W$ quantum memories. The memories can be processed in parallel, and a BSM can be performed on any pair of quantum memories within each repeater \cite{shi2020concurrent}.  

The protocol starts with the sender who tries to prepare $W$ EPR pairs and sends one end of each EPR pair through the optical link to the first repeater on the path to the receiver. Upon receiving the qubits from the sender, the first repeater sends an acknowledgment signal to the sender (which contains the indices of qubits successfully received), prepares $W$ EPR pairs, and sends one qubit of each EPR pair to the second repeater on the path. The second repeater similarly sends an acknowledgment signal to the first repeater and sends the first qubit of each  $W$ prepared EPR pair to the third repeater. As soon as the first repeater receives the acknowledgment signal from the second receiver (which contains the indices of successful EPR pairs between the first and second repeaters), the first repeater makes BSM and releases the outcomes to the neighboring nodes. Then, the second repeater performs BSM after receiving the acknowledgment signal from the third repeater and the outcome of BSM in the first repeater. This process continues with the next repeaters until we reach the receiver on the other end. Figure~\ref{fig:Path_W} shows an instance of our protocol for a path with $W=2$. The solid (dashed) lines indicate successful (failed) EPR trials and BSM is performed on successful links to generate an end-to-end EPR pair.

We assume a hard cut-off for the coherence time of quantum memories beyond which the memory is erased.
As a result, end-to-end entanglement may not be established due to the short coherence time of the repeater memories or those of end nodes. We use $N_{max}$ to indicate the number of repeater budget and $W_E$ as the number of memories at end nodes. $T_{RM}$ and $T_{EM}$ indicate the memory coherence time for repeaters and end nodes respectivly.
The optical link propagation time is $\tau_l(l_{uv}) = l_{uv}/c$ where $l_{uv}$ is the graph distance between the two nodes $u$ and $v$. \textcolor{black}{For simplicity, we will use the index of a link such as $i$ instead of $uv$ in the subscript of $l$ in some places of the paper.} The time required to generate an end-to-end entanglement is denoted by $\tau_{e2e}(.)$ which includes the classical messages exchange between consecutive repeaters on a given path as explained above.


\begin{table}
    \begin{tabularx}{\columnwidth}{c X }
      \toprule
$U$ & \textcolor{black}{Set of end nodes}\\
[0.18cm]
$Q$ & Set of user pairs\\
[0.18cm]
$R$ & Set of potential locations for repeaters\\
[0.18cm]
          {$N_\text{max}$}  & Number of repeaters budget   \\
        [0.18cm]
        $D_u$& Number of memories at repeater $u$\\
        [0.18cm]
    $D$ & Upper bound on repeaters memory capacity\\
    [0.18cm]
        { $W_E$} & Number of memories at end nodes \\
        [0.18cm]
        {$T_{RM}$}  & {Memory coherence time for repeaters}   \\
         [0.18cm]
              {$T_{EM}$}  & {Memory coherence time for end nodes}   \\
          [0.18cm]
        $q_{s}$ & {Success probability of Bell-state measurement}
          \\
          [0.18cm]
          $p_l(l)$ & {Transmission probability of link with length $l$}
          \\
          [0.18cm]
          $\tau_l(l)$ & {Returns the transmission delay on link with length $l$}
          \\
          [0.18cm]
          $\tau_{e2e}(p)$ & {Returns the delay time to deliver end-to-end ebit on path $p$}
          \\
\hline
      \bottomrule
    \end{tabularx}
    \caption{\protect\label{table:notaions} List of quantum network parameters. }
\end{table}
Consider a path with $h$ links (corresponding to $h-1$ repeaters) and width $W$, where the success probability of link-level EPR pair on the $i$-th link is $p_i=p_l (l_i)$ where $i = 1, 2, ..., h$ and
\begin{align}
    \label{eq:link-success}
    p_l (x) = 10^{-\alpha l_i}.
\end{align}
\textcolor{black}{Here, $l_i$ is the length of optical link $i$ (as an optical fiber) in km}, and $\alpha = 0.02$ is the signal attenuation rate in optical fiber (using $0.2$ dB/km at Telecomm wavelength).
The average end-to-end ebit generation rate (or throughput in short)
 can then be computed using a recurrence relation proposed in ~\cite{shi2020concurrent} (see Appendix~\ref{app:qcast} and equation (\ref{eq:recursive_equation2}) for details). The recurrence relation leads to nonlinear equations characterizing the end-to-end rate which makes the objective function for our optimization problem nonlinear.
Although there are ways to make our problem linear (as we explain in Sec.~\ref{sec:opt}), solving the recurrence relation for each path is time-consuming, especially for long paths (with larger $h$). This can easily add up to increase the overall time to compute the optimal solution. As a result, we approximate the average throughput by 
\begin{align}
\label{eq:e2e-rate}
    R_{e2e}(p,W) = q_{s}^{h-1} \cdot W \cdot p_\text{min},
\end{align}
where $p_\text{min}=\min (p_1,p_2,\cdots,p_h)$ is the minimum link-level success probability on the path. As we explain in Appendix~\ref{app:qcast}, this approximation is valid in the regime where $W p_\text{min}\gg 1$. \textcolor{black}{Here, we denote the entanglement swapping success probability by $q_s\leq 1$. Physically, finite success probability arises because direct two-qubit entangling gates (necessary for BSM) on most promising candidates for quantum memories based on ensemble of atoms~\cite{namazi2017ultralow}/ions~\cite{pettit2023perspective} or individual vacancy color centers~\cite{wang2023field} are either impossible or challenging; therefore, photon-mediated gates based on linear optics are often employed. The minimum success probability is $q_s=\frac{1}{2}$, which can be further increased by the so-called boosted fusion gates~\cite{grice2011arbitrarily,ewert2014efficient}. Throughout our paper, we set $q_s=\frac{1}{2}$ unless stated otherwise (in Section~\ref{sec:eff_swap_prob} we study the effect of $q_s$).} For reference, the end-to-end ebit rate associated with temporal or frequency multiplexing in a multimode memory corresponding to a path with $W=1$ is given by
\begin{align}
\label{eq:e2e-rate-tf-mx}
    R_{e2e} = q_{s}^{h-1} \prod_{i=1}^h \left(1-(1-p_i)^M \right),\end{align}
where $M$ is the multiplexing factor (see e.g.,~\cite{azuma2022quantum,sinclair2014spectral} and references therein).

\textcolor{black}{The quality of an end-to-end ebit is often characterized by its fidelity. We assume link-level ebits are in the form of Werner states~\cite{werner1989quantum} as in  }
\begin{align}
    \hat \rho_\text{ebit} =  \mu_L \ket{\psi^+}\bra{\psi^+} 
    + (1-\mu_L)\frac{\hat {{1}}}{4},
\end{align}
\textcolor{black}{the fidelity of which is given by $F_L = (1+3\mu_L)/4$. As we see from the above equation, Werner states are a result of applying a depolarizing channel to a perfect Bell state $\ket{\psi^+} = (\ket{01}+\ket{10})/\sqrt{2}$. These states are a common choice in the literature to describe a generic (most pessimistic but hardware agnostic) noisy Bell state of two qubits and were originally used in Refs.~\cite{briegel1998quantum, bennett1996purification} to showcase the utility of purification schemes to produce a high fidelity ebit out of multiple very noisy ebits. A Werner state is parameterized by a single parameter $\mu_L$, where a perfect state $\mu_L=1$ gives $F_L=1$ and a fully depolarized state $\mu_L=0$ correspond to $F_L = 1/4$.
Another feature of a Werner state is that it greatly simplifies calculations in the following sense: the post-swap state of a sequence of noisy ebits in the form of Werner states on a repeater chain results in an end-to-end ebit in the form of another Werner state parameterized by $\mu_h = \mu_L^h$~\cite{briegel1998quantum, bennett1996purification}. More generally, if the BSM process is noisy 
the end-to-end fidelity is given by~\cite{briegel1998quantum}}
\begin{align}
    \label{eq:e2e-fidelity}
     F_{e2e}= \frac{1}{4} + \frac{3}{4}{(\frac{P_2(4\eta^2-1)}{3})}^{h-1} {(\frac{4F_{\text{L}}-1}{3})}^{h},
\end{align}
\textcolor{black}{where the BSM are considered noisy due to the controlled-not gate needed for BSM) modeled by a two-qubit depolarizing channel parametrized by gate fidelity $P_2$ and a final measurement characterized by measurement fidelity $\eta$. For simplicity, we assume the BSM process is not noisy $P_2=\eta=1$ in our numerical experiments. }

Regarding scheduling of link-level entanglement generation,
one may consider a parallel protocol where the main difference with the above sequential protocol is that all repeaters on a path start generating link-level entanglement simultaneously. Such a parallel protocol gives the same end-to-end success probability as Eq.~(\ref{eq:e2e-rate}) while it can reduce $\tau_{e2e}$, ultimately leading to larger ebit rate per unit time, $R_{e2e}/\tau_{e2e}$. 
This is however 
at the expense of longer run times for repeater memories since regardless of the link-level synchronization protocol the BSMs must be performed sequentially from sender to receiver. In other words, a given repeater needs to know the indices of successful BSMs in previous steps to determine which quantum memories of theirs are entangled with the sender's memories. 
We imagine a future quantum network to have lower-quality memories (with shorter coherence time) inside repeaters (i.e., network core) and high-quality memories (with longer coherence time or possibly fault-tolerant) at the end users (i.e., network edge). Therefore, we adopt the sequential protocol as it imposes a less strict requirement on the coherence time of repeater memories. To increase the end-to-end ebit rate per unit time, we can increase $R_{e2e}$ by increasing the path width $W$ (c.f.~Eq.~(\ref{eq:e2e-rate})).

\subsection{Objective}

The objective of our network planning optimization problem is to maximize the aggregate utility of the set $Q$ of user pairs. The quantum utility function of a user pair is defined as
\begin{align}
\label{eq:utility}
    U(R_{e2e},F_{e2e})=\log_2 (R_{e2e}\cdot f(F_{e2e})),
\end{align}
in terms of the end-to-end rate $R_{e2e}$ and fidelity $F_{e2e}$ of the EPR pairs delivered to them. This is the \textit{negativity} quantity proposed in \cite{vidal2002computable} to quantify the degree of entanglement in composite systems. Other possible utility functions are distillable entanglement \cite{rains2001semidefinite} and secret key rate \cite{bennett2014quantum} as explored in \cite{vardoyan2022quantum}. Here, the functional form of $f(F_{e2e})$ depends on the application and takes different forms for computing~\cite{lee2022quantum} and networking, or secret sharing~\cite{bennett2014quantum}. In this paper, we use the following formula based on entanglement negativity~\cite{vidal2002computable},
\begin{align}
\label{eq:negativity}
    f(F) = F - \frac{1}{2}\ ,
\end{align}
as a proxy for the quality of the end-to-end ebits, since it is an upper bound on the distillable entanglement~\cite{plenio2005logarithmic}. The utility function based on \textit{negativity} is preferred as it is
concave and one can use convex optimization techniques to efficiently find the optimal value \cite{vardoyan2022quantum}.

\subsection{Planning output}

The output of the optimization problem provides four results: (1) the number of repeaters to be used, (2) where to place them in the network, (3) the paths for each user pair, and (4) the assigned quantum memories at the repeaters to different paths. 

\subsubsection{Definition of a path}
\label{section:path_definition}
A path is a sequence of repeater locations. Two consecutive locations on the path are not necessarily two consecutive nodes on the actual graph. We assume there is a direct physical link between each two locations with the length of the shortest path between them. This allows us to have more than one path between two end nodes even on a repeater chain. For example, in Figure \ref{fig:Path_W}, we can have paths $[1,5],[1,3,5] ,[1,2,3,4,5]$ and $[1,3,4,5]$ etc. The path $[1,5]$ means none of the locations have been chosen to have a repeater and that means no repeater is used. In this figure, we have the path $[1,3,4,5]$ which means there is a direct physical link between node $1$ and node $3$. The length of this link is the summation of the length of the link that connects node $1$ to $2$ and the link that connects node $2$ to node $3$.

Figure \ref{fig:Path_W} illustrates an example of the output of our optimization. It is a linear chain with nodes 1 and 5 as users and 3 potential places for repeater placement: nodes 2, 3, and 4. The optimal solutions places two repeaters at nodes 3 and 4. Note that node 2 is grayed out which means this node will not be used as a repeater but rather an optical switch providing an optical link between 1 and 3.  In this example, since there is only one user pair, the optimal solution is to assign both memories to this user pair to maximize the end-to-end ebit rate.

Before closing this section, let us make a few remarks on related previous work. A similar idea for network planning but with one multi-mode memory per channel (c.f., Eq.~(\ref{eq:e2e-rate-tf-mx})) has been proposed in \cite{rabbie2022designing}. Our work is similar to their work as we also use the preexisting infrastructure for network planning. However, our goal is to maximize the network utility, which favors short paths with few repeaters.
In addition, we consider a different type of quantum memory scheme using spatial multiplexing (c.f., Eq.~(\ref{eq:e2e-rate})) and analyze the effect of the finite coherence time of quantum memories. In contrast to Ref.~\cite{rabbie2022designing} which uses equally-distanced repeaters to estimate the end-to-end entanglement rate of a given path (regardless of the repeater positions), we evaluate the entanglement rate for each path specifically based on the exact location of the repeaters. 

\section{Optimization problem} 
\label{sec:opt}

In this section, we present two equivalent ways of formulating the quantum network planning problem and discuss how we turn them into integer (binary) linear programs. \textcolor{black}{We first explain our path-based problem formulation and its constraints ( $\S$\ref{subsection:path_based}) then we explain our link-based problem formulation with its constraints ($\S$\ref{subsection:link_based}).}

\begin{table}
    \begin{tabularx}{\columnwidth}{c X }
      \toprule
         {$P^{q}$}  & {Set of all  paths for user pair $q$ in $G$}  \\
    [0.18cm]
            $K$& Number of allowed paths for each user pair\\
        [0.18cm]
        $r_u$ & Indicates whether
node $u \in R$  is used as a repeater node or not\\
        [0.18cm]
    $x^q_p$ & Indicates whether
path $p$  is used for user pair $q$ or not\\
       [0.18cm]
$w^q_{p}$ & Width of path $p$ for user pair $q$\\
        [0.18cm]
\hline
      \bottomrule
    \end{tabularx}
    \caption{\protect\label{table:notaions-path-based} List of variables used in the path-based formulation. }
\end{table}

\subsection{Path-based formulation}
\label{subsection:path_based}
Equations (\ref{problem:exhaustive_search_utility_maximization2})-(\ref{cons:variable_constraint3}) define our path-based network planning optimization problem. \textcolor{black}{We assume each user pair $q$ can use up to $K\geq 1$ paths from the set of $P^q$ paths in the network where $P_q$ is the set of all paths available for user pair $q$. In our simulations in the following sections, we set $K=1$ for simplicity.}. Let $r_u \in \{0,1\}$ denote whether node $u$ is used as a repeater node or not.  Given path $p$, the end-to-end throughput $R_{e2e}(.)$ and fidelity $F_{e2e}(.)$ are computed using Eqs.~(\ref{eq:e2e-rate}) and (\ref{eq:e2e-fidelity}), respectively.
Decision variables are $x_p^q$ and  $w_p^q$ which indicate the path that should be used for user pair $q$ and the width of path $p$, e.g., the number of memories to deploy on the entire path $p$ (width of path $p$) for source-destination pair $q$. This in turn implies which nodes to be used as repeaters where $r_u=1$. 

\myparab{Problem 1 (path-based problem formulation)}
\label{problems:problem1}
\begin{align} 
\max_{r_u,x^{q}_p,w^q_p}  & \sum_{\substack{q \in Q }} \sum_{p\in P^q}U(R_{e2e}(p,w^q_p),F_{e2e}(p))x^{q}_p 
\label{problem:exhaustive_search_utility_maximization2}
\\
\text{s.t.}& 
\nonumber\\
& \sum_{\substack{q \in Q \\
p \in P^q |u \in p
}
} w^q_p\cdot x^{q}_{p}  \leq D_u r_u
\label{cons:repeater_memory_constraint} \quad \quad \quad  \quad \quad \quad \quad \forall u\in R  
\\
& \sum_{\substack{p \in P^q
}
}  x^{q}_{p}  \leq K \quad \quad \quad  \quad \quad \quad \quad \quad \forall{q\in Q}
\label{cons:using_at_most_k_paths}
\\
& \textcolor{black}{\sum_{\substack{p \in P^q
}
}  w^q_p\cdot x^{q}_{p}  \leq W_E} \quad \quad \quad \quad \quad \quad \quad \forall{q \in Q}
\label{cons:using_at_most_end_M_memories}
\\
& \sum_{u \in R} r_u  \leq { N_\text{max}}
\label{cons:using_at_most_N_repeaters} \quad \\
& { 2\tau_l(l_e) \cdot x^q_p  \leq T_{RM} \quad \quad  \forall{q \in Q} ,  \forall{p \in P^q}, \forall{e \in p}}
\label{cons:repeater_life_time_constraint} \\
&  {\tau_{e2e}(p)}\cdot x^q_p  \leq T_{EM} \quad \quad \quad \quad  \quad \forall{q \in Q} ,  \forall{p \in P^q}
\label{cons:end_node_life_time_constraint} \quad \\
\quad & x^{q}_{p}  \in \{0,1\} , \quad \quad \quad \quad \quad \quad \quad  \quad \forall q \in Q,  p \in P^{q}
\\
\quad & r_{u} \in \{ 0,1\} , \quad \quad \quad \quad \quad \quad \quad \quad \quad \quad \quad \forall u \in R \\
\quad & w^q_{p} \in \{ 1,2,3,...min(D,W_E)\}, \quad  \forall q \in Q ,  p \in P^q 
\label{cons:variable_constraint3} 
\end{align}
Constraint~(\ref{cons:repeater_memory_constraint}) ensures at most $D_u$ memories at network node $u$ which is selected as a repeater. Constraint ~(\ref{cons:using_at_most_k_paths}) ensures that only $K$ paths be used for each user pair. Constraint~(\ref{cons:using_at_most_end_M_memories}) enforces the memory limit of end nodes and constraint~(\ref{cons:using_at_most_N_repeaters}) ensures that we use at most $N_\text{max}$ repeaters in the network.
Constraint~(\ref{cons:repeater_life_time_constraint}) ensures that for each path decided to be used in the network and for all optical links on that path, the time required to generate entanglement and receive the acknowledgment signal for ebit generation must be less than or equal to the memory coherence time of the repeaters.
Constraint~(\ref{cons:end_node_life_time_constraint}) ensures the time required for end-to-end entanglement generation to be less than the end-node memory coherence time for a selected path. $D$ is the maximum number of memories for all deployed repeaters.


The above problem formulation has two drawbacks. First, the objective function as defined in Sec.~\ref{sec:protocol} is nonlinear which makes the problem an integer non-linear program. { We also have a product of two decision variables in memory constraints~(\ref{cons:repeater_memory_constraint}) and (\ref{cons:using_at_most_end_M_memories}).} 
We resolve this issue by enumerating all the versions of each path (including possible values for the path width) and computing the nonlinear utility function. 
Second, it is not practical to enumerate all paths for large networks { (which implies $|Q| \min(D,W_E) |R|!$ variables for $x_p^q$)
since this scales exponentially with the number of nodes $|R|$.} In order to resolve this issue, we note that we may not need to enumerate all the paths in the network to obtain either the optimal or a near-optimal solution. {
Instead, we use the algorithm proposed in \cite{yen1971finding} to find the first $k$ shortest paths and run our path-based optimization algorithm (\ref{problem:exhaustive_search_utility_maximization2}) on the reduced set.
As we show in the evaluation section, we can reach the optimal solution by limiting the number of decision variables to $|Q| \min(D,W_E) k$ where $k=1000-4000$ for random networks with $|R|\leq 50$ (Appendix~\ref{app:k-paths}) and dumbbell topology with $|R|\leq 10$ (Sec.~\ref{sec:dumbbell}).}
Alternatively, as we explain next, one can formulate a link-based version of the same problem {where the number of decision variables scales polynomially with the number of network nodes}.

\subsection{Link-based formulation}
\label{subsection:link_based}
%
%

Here, we present a link-based formulation of the quantum network utility maximization problem. For each user pair $q=(s,t)$, we define an array of binary variables $x^{q}_{uv}$ associated with each directed link $(u,v) \in {\cal E}_q$ where the set of links is defined as
\begin{align}
{\cal E}_q = \{ (u,v)| u \in R \cup \{s\}, v \in R \cup \{t\}, u\neq v \}.
\end{align}
An end-to-end path is described by a subset of $x^{q}_{uv}$ variables that are non-zero. 
Constraint~(\ref{cons:avoiding_loops}) is the flow continuity equation (similar to the maximum flow problem \cite{goldberg1988new}) to ensure that there is a directed path between the sender $s$ and receiver $t$.
For instance, the solution in Figure~\ref{fig:Path_W} for $q=(1,5)$ corresponds to $x_{13}^q=x_{34}^q=x_{45}^q=1$ with other entries being zero. Functions $s()$ and $t()$ return the sender and receiver of user pair $q$ respectively.

\begin{table}
    \begin{tabularx}{\columnwidth}{c X }
      \toprule 
$x_{uv}^{q,w}$
 & Indicates whether
the link $(u, v)$ with width $w$ is used as part of a path for $q$ user pair or not\\
        [0.18cm]
    $d_q$ & Longest link for a path connecting user pair $q=(s,t) \in Q$\\
       [0.18cm]
    $\beta_{q,w}$ & Indicates whether
path with width $w$  is used for user pair $q$ or not\\
       [0.18cm]
    $s(q)$ & Indicates the sender node of user pair $q$\\
       [0.18cm]
    $t(q)$ & Indicates the receiver node of user pair $q$\\
       [0.18cm]
\hline
      \bottomrule
    \end{tabularx}
    \caption{\protect\label{table:notaions-link-based} List of variables used in the link-based formulation. }
\end{table}

The objective function (\ref{problem:linear_link_based_utility_maximization}) is the aggregate utility (\ref{eq:utility}) where we rewrite the end-to-end ebit rate Eq.~(\ref{eq:e2e-rate}) using the decision variables $x_{u,v}^{q,w}$. We note that the utility function defined in (\ref{eq:utility}) does not necessarily favor more repeaters. This is not only because the end-to-end fidelity (\ref{eq:e2e-fidelity}) decreases as we add more links (or increase $h$) but also because the overall swapping success probability decreases in the end-to-end rate (\ref{eq:e2e-rate}). Therefore, even if we set $F_L=1$ (which implies $F_{e2e}=1$) and neglect the impact of fidelity the optimal solution may only use a fraction of potential locations for repeaters. Based on this observation, we omit the fidelity from the utility function so that we can reduce our link-based formulation to an integer linear programming. We recall that the role of the fidelity term is to penalize overusing repeaters, and we still have another term, namely, the overall swap success probability in the end-to-end rate (\ref{eq:e2e-rate}) to enforce that. Since the dependence of the utility function on path width $w$ is nonlinear (i.e., $\log_2 w$), we cannot use $w$ as a decision variable and maintain a linear programming problem. Hence, we introduce $W_E$ copies of $x_{u,v}^{q,w}$ and include $w=1,\cdots,W_E$ as a superscript and auxiliary variable $\beta_{q,w}$ defined in (\ref{cons:auxiliary_constraint1}) is an array of size $W_E$ where the only non-zero element determines which value of $w$ is used.

\myparab{Problem 2 (link-based problem formulation)}
\label{problems:problem2}
\begin{align} 
\max & \sum_{q\in Q} [ \log_2 (q_s) \left(\sum_{w,(u,v)\in {\cal E}_{q}} x^{q,w}_{uv}-1 \right) \nonumber+ \\ 
& \sum_{w} \beta_{q,w} \log_2 (w) -\alpha_2 d_q ]
\label{problem:linear_link_based_utility_maximization}
\\
\text{s.t.}\quad &  \sum_{v,w} x^{q,w}_{uv} - x^{q,w}_{vu} =
\begin{cases}
1, \quad \quad  \quad \quad\text{if} \quad \textcolor{black}{u=s(q)}\\
- 1, \quad \quad \ \quad \text{if} \quad \textcolor{black}{u=t(q)}\\
0, \quad \quad  \quad \quad \text{if} \quad u\in R 
\end{cases}\label{cons:avoiding_loops} \nonumber \\
\quad \quad \quad  & \quad \quad  \quad \quad  \quad \quad \quad \quad \quad \textcolor{black}{\forall{q\in Q}} \quad \& \quad \textcolor{black}{\forall{u \in R}} \\
&\sum_{w} x_{uv}^{q,w} \leq 1 \hspace{1.4cm} \forall (u,v) \in {\cal E}_q, \forall q \in {Q} \label{cons:using_each_edge_once} \\
&\beta_{q,w} = x^{q,w}_{st} + \sum_v x^{q,w}_{sv} \hspace{1.1cm} \forall w,\ \forall q \in {Q}\label{cons:auxiliary_constraint1} \\
&d_q \geq l_{uv} x_{uv}^{q,w} \hspace{1.3cm} \forall (u,v) \in {\cal E}_q, \forall q \in {Q} \label{cons:auxiliary_constraint2} \\
&\sum_{w,q,v} w  x_{uv}^{q,w} \leq D_u r_u \hspace{2.4cm} \forall u\in {R} \label{cons:repetaer_capacity}\\
&\sum_{w,v} w  x_{sv}^{q,w} \leq W_E \hspace{2.6cm} \forall q \in {Q} \label{cons:end_capacity}\\
&\sum_u r_{u} \leq {N_{\max}} \hspace{2.5cm} \forall u\in {R}
\label{eq:max-repeater} \\
& 2\tau_l(l_{uv}) x_{uv}^{q,w} \leq T_{RM} \hspace{0.6cm} \forall w, \forall (u,v) \in {\cal E}_q, \forall q \in {Q} 
\label{cons:repeater_Tcoh}\\
& 3 \sum_{(u,v)} \tau_{l}(l_{uv}) x_{uv}^{q,w} \leq T_{EM} \hspace{1.1cm} \forall w, \forall q \in {Q} 
\label{cons:end_Tcoh}\\
&x_{uv}^{q,w} \in \{0, 1\} \hspace{2.1cm} \forall w,\ \forall (u,v) \in {\cal E}_q \\
&r_{u} \in \{0, 1\} \hspace{3.7cm} \forall u \in {R} \\
&\beta_{q,w} \in \{0, 1\} \hspace{2.6cm} \forall w,\ \forall q \in {Q} 
\end{align}

We impose constraint~(\ref{cons:using_each_edge_once}) to ensure that only one path (out of $W_E$) will be chosen.
The summand in the objective is $\log_2 R_{e2e}$ as defined in (\ref{eq:e2e-rate}) and can be understood as follows: the first term is $\log_2 q_s^{h-1}$ where we rewrite the number of active links as a sum over all entries of $x_{u,v}^{q,w}$. The second term accounts for which value of path width is used and the last term is the minimum success probability (\ref{eq:link-success}) on a path after taking the logarithm, i.e., $\log_2 p_\text{min} = \log_2 10^{-\alpha d_q} = -\alpha_2 d_q$ where $\alpha_2 =\alpha\log_2 10$, and $d_q$ gives the longest link on the path (calculated via constraint~(\ref{cons:auxiliary_constraint2})).

Let us now discuss the remaining constraints in the link-based formulation.
 Constraints~(\ref{cons:repetaer_capacity}), (\ref{cons:end_capacity}), and (\ref{eq:max-repeater}) are identical to 
constraints~(\ref{cons:repeater_memory_constraint}),
(\ref{cons:using_at_most_end_M_memories}), and
(\ref{cons:using_at_most_N_repeaters}) in the path-based formulation, which impose repeater memory, end-user memory, and a maximum number of repeater 
 constraints, respectively.
Constraint~(\ref{cons:repeater_Tcoh}) is analogous to (\ref{cons:repeater_life_time_constraint}) in the path-based approach and does not allow links where the signal round trip takes longer than the repeater memory coherence time. Lastly, constraint~(\ref{cons:end_Tcoh}) ensures the end-to-end entanglement distribution process does not take longer than the memory coherence time at the end users. We note that (\ref{cons:end_Tcoh}) has the benefit of being a linear constraint at the cost of being more stringent than (\ref{cons:end_node_life_time_constraint}) in the path-based formalism.

We note that the link-based formulation reduces the problem size (i.e., number of entries in $x_{uv}^{q,w}$) to $|Q| \min(W_E,D)  [|R|(|R|+1)+1]$  which is significantly smaller than the path-based approach.


Although we do not use entanglement distribution protocols based on frequency or time multiplexing in our simulations, it is worth noting that the utility function associated with the rate in this case (\ref{eq:e2e-rate-tf-mx}) can also be written as a linear function,
\begin{align}
    \log_2 R_{e2e} = \log_2 (q_s) \left(\sum_{(u,v)\in {\cal E}_{q}} x^q_{uv}-1 \right) \nonumber+ \\ \sum_{(u,v)\in {\cal E}_{q}} {x^q_{uv}} \log_2\left(1-(1-p_l(l_{uv}))^M \right),
\end{align}
where superscript $w$ is dropped since this multiplexing scheme assumes one memory per channel.

\subsection{Scale invariance and equivalence between the two formulations}

One way to prove the equivalence of the two formulations, \textcolor{black}{under the assumption that the set of paths for each user pair in the path-based formulation consists of all possible paths and that $K$ and the link level fidelity is $1.0$,} is to show that an optimal solution in each of the formulations maps to a valid solution in the other formulation~\cite{chakraborty2020entanglement,rabbie2022designing}. It is easy to see why. Suppose the optimal utility function for path-based and link-based schemes are denoted as $U_p$ and $U_l$. 
An optimal path $p$ in the path-based formulation consists of some links connecting the two end users, which in the link-based formulation corresponds to setting those entries in $x_{uv}^{q,w}$ one and keeping the rest zero.
The path $p$ is then a valid solution to the link-based formulation since all the constraints in either formulation are equivalents. Hence, we have $U_p \leq U_l$. Similarly, the optimal set of activated links given in terms of the array $x_{uv}^{q,w}$ can be viewed as a path where only $u, v$ nodes with $x_{uv}^{q,w}=1$ are being used. Therefore, we can write $U_l \leq U_p$. The two inequalities have to be satisfied simultaneously which implies $U_l=U_p$, i.e., the two optimal solutions are identical.

We note that either formulation of the problem enjoys a scale invariance property as follows:
The problem does not change as we rescale repeater capacity $D \to \lambda D$, end user capacity $W_E \to \lambda W_E$, and $w \to \lambda w$
by a scaling factor $\lambda$. This is because the network capacity constraints~(\ref{cons:repetaer_capacity}), (\ref{cons:end_capacity}), (\ref{cons:repeater_memory_constraint}),
and (\ref{cons:using_at_most_end_M_memories}) remain the same after the rescaling and the objective function is shifted by a constant $|Q| \log_2 \lambda$ (which can be removed). Therefore, the optimal solution remains the same and the optimal number of memories for each pair scales the same way $w_\text{opt} \to \lambda w^q_\text{opt}$. This means that only relative ratios are relevant, i.e., which portion of repeater memories $\frac{w^q_\text{opt}}{D}$ are assigned to user pair $q$. For instance, if we have two user pairs and the optimal solution for $D=10$ is $w^{q_1}_\text{opt}=w^{q_2}_\text{opt}=5$, it means that if we solve the problem for $D=1000$, then we simply have $w^{q_1}_\text{opt}=w^{q_2}_\text{opt}=500$. 

The scale invariance is an important property of our formulation for the following reason: Suppose we run an optimization problem for a small value of repeater capacity $D=10$ so that the problem size is small and manageable and find the optimal path with $w^q_\text{opt}=3$ to have the longest link of length $100$km. This solution violates our approximation for the end-to-end ebit rate (\ref{eq:e2e-rate}) which requires $w^q_\text{opt} p_\text{min} \gg 1$ while we have $w^q_\text{opt} p_\text{min}=0.03$. Thanks to the scale invariance property, we can say our solution, $\frac{w^q_\text{opt}}{D}=0.3$, is still valid for $D \gtrsim 1000$ which implies the minimum number of memories to be $w^q_\text{opt} = 300$. We use this fact when we run quantum network planning for the ESnet.

\section{Evaluation}
\label{experiment}

\begin{figure*}
\centering
\includegraphics[scale=0.65]{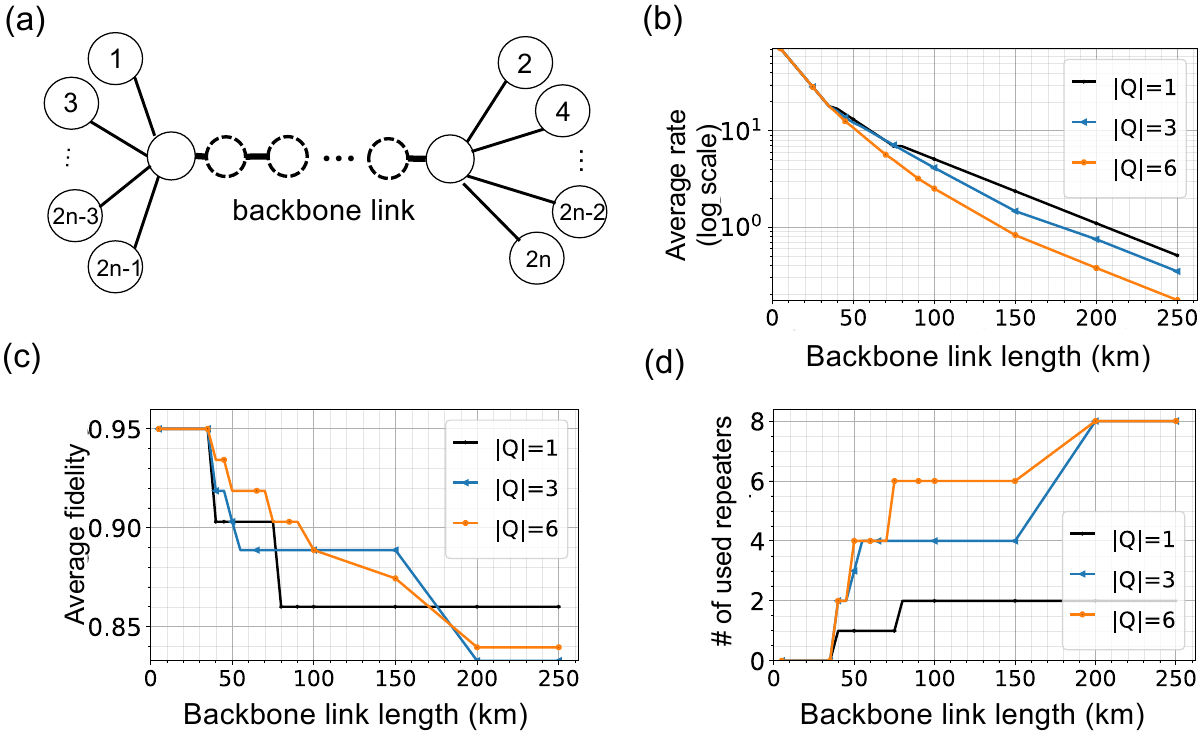}

\caption{\textcolor{black}{(a) Dumbbell topology with $|Q|=n$ user pairs and 10 potential places for repeater placement (dashed circles)}, (b) optimal
ebit rate per user pair, (c) optimal end-to-end fidelity, and (d) number of used repeaters as a function of the backbone link length.}
\label{fig:dumbbell_topology}
\end{figure*}

In this section, we report insights from our experiments. We use a synthetic (dumbbell-shape geometry) and two real-world topologies (SURFnet and ESnet) as our physical topologies. We use IBM CPLEX solver to solve the linearized version of our optimization problem (~\ref{problem:exhaustive_search_utility_maximization2}) and (~\ref{problem:linear_link_based_utility_maximization}). We assume the entanglement swapping success probability, $q_s$, is $\frac{1}{2}$ unless it is mentioned otherwise. In addition, the fidelity of a link-level ebit is $0.95$ unless stated otherwise and we assume the maximum number of memories for all deployed repeaters is identical, i.e., $D_u=D$.



\begin{figure}
\centering
\includegraphics[scale=0.43]{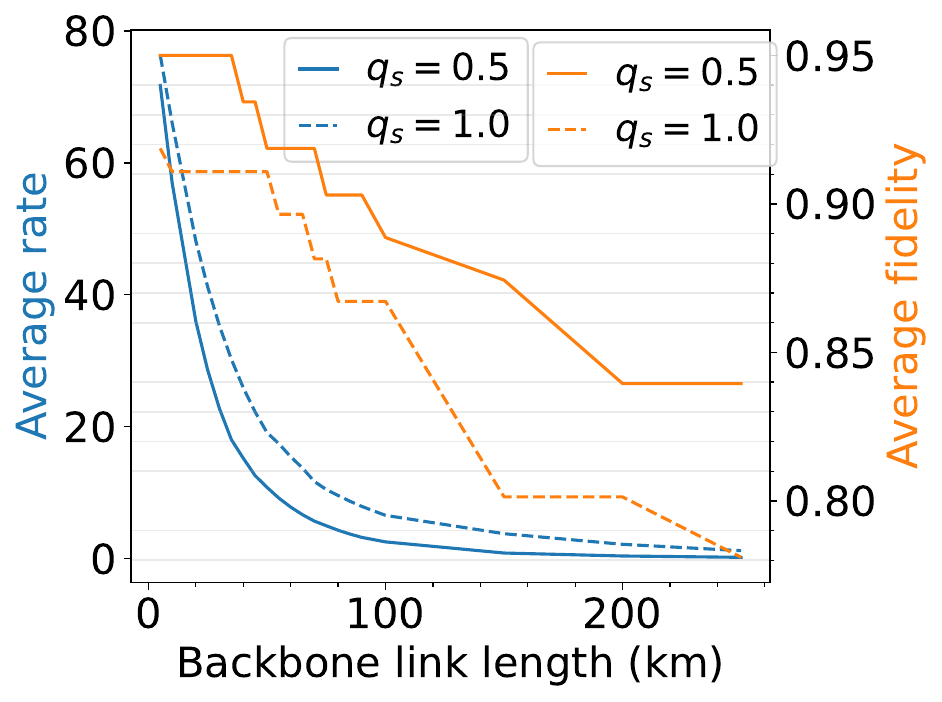}
\caption{\textcolor{black}{The effect of $q_s$ on the average end-to-end rate and average fidelity for $|Q|=6$ user pairs in our synthetic topology (figure \ref{fig:dumbbell_topology}.a)
}}
\label{fig:dumbbell_topology_affect_of_q_s}
\end{figure}

\subsection{Synthetic topology}
\label{sec:dumbbell}

Our synthetic topology is a dumbbell-shape geometry shown in Figure \ref{fig:dumbbell_topology}(a). In this topology, there are $n$ user pairs connected through a backbone link. In Figure \ref{fig:dumbbell_topology}(a) node $1$ is paired with node $2$, node $3$ is paired with node $4$, and so on. The length of the link connecting each node to the closer end of the backbone link is $1$km. We vary the length of the backbone link in this experiment. 

We use our path-based formulation (\ref{problem:exhaustive_search_utility_maximization2}) in this experiment as it includes end-to-end fidelity in the utility function. As mentioned in the previous section, here we use the $k$-shortest paths algorithm ~\cite{yen1971finding}
and consider $|P^q|=4,000$ paths for each user pair and $K=1$. Note that we can have more than one path in a repeater chain based on our definition of a path in section ~\ref{section:path_definition}.
We set $W_E=D= 100$,
and do not impose constraints on the memory coherence time of repeaters or end nodes in this experiment.


\subsubsection{Utility vs. Distance between repeaters}

The first experiment demonstrates how the utilities of user pairs change as we increase distances between potential places for repeaters. For this, we consider $|R|=10$ locations for repeaters at equal distances $L/(|R|+1)$ along the backbone link with length $L$ as shown in Figure \ref{fig:dumbbell_topology}(a). The distance between the potential repeater locations is increased uniformly by increasing $L$, and we solve the optimization problem for each value of $L$. 

Figure \ref{fig:dumbbell_topology}(b) shows how the optimal end-to-end entanglement rate for each user pair varies as we increase the backbone link distance. When there is only one user pair in the network, all memories available in the repeaters are assigned to that user pair and it receives a high rate compared to cases where we have more than one user pair. In the presence of more user pairs, the pairs share repeaters and receive fewer memories to maximize the aggregate utility function Eq.~(\ref{eq:utility}).


\begin{figure*}
    \centering
    \includegraphics[scale=0.74]{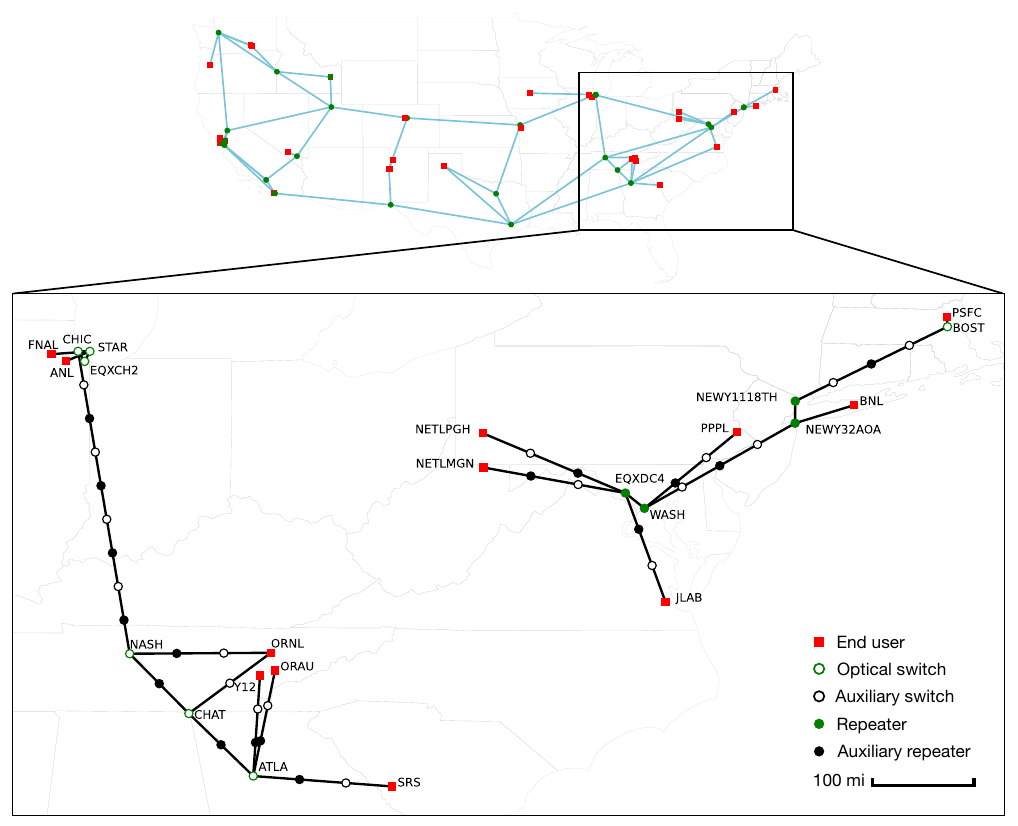}
    \caption{\protect\label{fig:esnet} Optimal locations of repeaters for the augmented subgraph of the ESnet including nodes in the East Coast and the Midwest. The black circles (open and filled) denote the auxiliary nodes placed to make the longest link $100$km long. The optimization solution is shown as filled circles which indicate the locations of nodes turned into repeaters while open circles are not used. Some end nodes are shifted to improve readability.}
\end{figure*}

\subsubsection{Fidelity/Number of used repeaters vs. Distance between repeaters}
Figures \ref{fig:dumbbell_topology}(c) and \ref{fig:dumbbell_topology}(d) show the optimal end-to-end fidelity and the number of repeaters used in the network out of our $10$ repeater budget as a function of backbone link length. There is { an inverse} correlation between the number of repeaters and the average end-to-end user pairs fidelity. This is expected based on Eq.~(\ref{eq:e2e-fidelity}) as end-to-end fidelity on a path decreases as more repeaters are used. When the backbone link length is small (less than $40$km), no repeaters are used and there will be a direct link between the end nodes. As we increase the backbone link length, link-level ebit generation success probability decays exponentially and more repeaters are \textcolor{black}{placed} to increase link-level generation success probabilities. As is evident from the plot, the optimal solution never utilizes all $10$ available repeater locations in the network.

\subsubsection{Effect of $q_s$}

\label{sec:eff_swap_prob}
\textcolor{black}{In the last experiment of this section, we explore the affect of $q_s$ on resource placement for network planning. For that, we plan our synthetic network for $|Q|=6$ user pairs for two different values of $q_s$ ($q_s=\frac{1}{2}$ and $q_s=1.0$), which is illustrated in Fig. \ref{fig:dumbbell_topology_affect_of_q_s} in terms of the end-to-end rate and fidelity. We observe that with $q_s=1.0$, the average delivered end-to-end rate is higher than the rate for $q_s=\frac{1}{2}$ but the average fidelity is lower due to usage of a higher number of repeaters. The reason is that when $q_s$ is $1.0$, swaps do not decrease the rate (and the utility) since adding more repeaters increases the success probability of link-level entanglement generation rate by decreasing the distance between repeaters by adding more repeaters on a path. However, according to Eq. (\ref{eq:e2e-fidelity}), adding more repeaters will decrease end-to-end fidelity when $q_s< 1$. For the specific utility function in Eq. (\ref{eq:utility}) that we have used here, increasing the rate by adding more repeaters appears to be more advantageous for maximizing the utility function than improving end-to-end fidelity.}



\subsection{ESnet topology}

In this experiment, we use the ESnet topology \cite{esnet} and examine how repeater placement on this network affects utility maximization.
We have derived the geographical locations of the nodes from \cite{esnet} and estimated link lengths in terms of their geodesic distances.
We focus on the East Coast and the {Midwest} shown in Figure \ref{fig:esnet} and consider three user pairs in each region.
The ESnet core and edge nodes are shown as red squares and green circles in the upper panel of this figure.
Since the original links are long (greater than few hundred kilometers), we have augmented the network graph by adding auxiliary nodes so that no link is longer than $2L_0$ (to be specified for each experiment). We achieve this in the following way: Given a link of length $\ell > 2L_0$, we place $n_\ell = \lfloor \frac{\ell}{L_0} \rfloor -1$ repeaters.

\begin{figure}
    \includegraphics[scale=0.53]{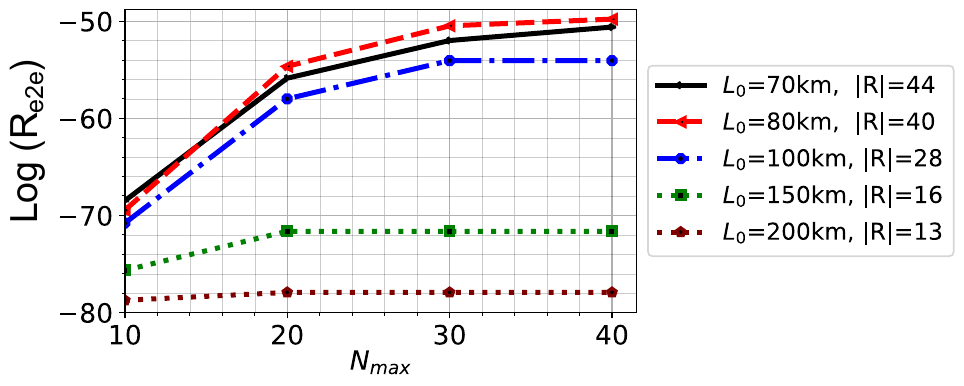}
    \caption{\protect\label{fig:esnet_max_distance} ESnet network planning on the  ESnet augmented network graph where we place additional repeaters to upper bound the maximum link length $L_0$ (see main text for details). The legend also shows the number of potential repeater locations after the augmentation. Here, we set the memory capacity of repeaters and end users to be $W_E=D=10$.}
\end{figure}

Figure \ref{fig:esnet_max_distance} shows how the maximum aggregate utility changes as we increase the repeater budget $N_\text{max}$ for different values of $L_0$.
We observe that increasing $N_\text{max}$ for a fixed value of $L_0$ initially improves utility but eventually saturates. 
This illustrates competition between repeater spacing and number of repeaters in the optimal solution (c.f.~Eq.~(\ref{eq:e2e-rate})) where adding more repeaters may increase link-level success probabilities but the end-to-end ebit rate decreases due to lower swap success probabilities. The fact that the saturation occurs for small values of $N_\text{max}$ depends on the details of the network topology.
We further see that decreasing $L_0$ from $200$km to $100$km increases the maximum aggregate utility but that the improvement diminishes as we further decrease $L_0$ below $100$km.

The lower panel of Figure~\ref{fig:esnet} shows an example of the optimal solution where the graph is augmented with repeater locations no more than $L_0 = 70$ km apart (added nodes are shown as filled and open black circles). With this value of $L_0$, we observe that the longest link has a length of $100$km. The result of optimization 
for the following set of user pairs are shown: (SRS, ORAU), (Y12, FNAL), (ORNL, ANL), (NETLPGH, PSFC), (NETLMGN, PPPL), and (BNL, JLAB).  
 Here, we use our link-based LP formulation (\ref{problem:linear_link_based_utility_maximization}) with link fidelity equal to one.
 In this case, after the augmentation, we have { $|R|=44$}, and we set $N_\text{max}=10$ for each region.
 We further show the individual paths for each user pair explicitly in Appendix~\ref{app:esnet}. 



\subsection{SURFnet topology}

In this experiment, we show how quantum memory coherence time and memory capacity at repeaters and end nodes affect the maximum quantum utility of the network. We use the SURFnet topology (Figure \ref{fig:SURFnet}(a)) in the next two experiments. We consider a set of 4 user pairs randomly selected in the network as the workload. In this experiment, we choose user pairs with distances in the range of $200$ and $250$ km from each other. For Figure \ref{fig:SURFnet}(b) we plot the average of $100$ different workloads and for Figure \ref{fig:SURFnet}(c) and \ref{fig:SURFnet}(d) we consider only one workload.  We assume we can use $N_\text{max}=10$ repeaters each with $D=100$ memories across the network. Each node in the SURFnet topology is a potential location for a repeater.

\subsubsection{Utility vs. Memory coherence time}

Figure \ref{fig:SURFnet}(b) shows the aggregate utility of the user pairs in SURFnet topology as a function of the memory coherence times of repeaters and end nodes. The $x$-axis is end node memory coherence time and the $y$-axis is repeater memory coherence time both in milliseconds ($m$s). When end node memory coherence time is less than $3.2$ $m$s, using repeaters with high-quality memories  (memories with a long coherence time for qubits) does not further increase utility. This is because end node memory coherence time does not support holding qubits for entanglement generation and receiving the heralding signal across any path (even the shortest path). In this experiment, we set the aggregate utility to $-50$ when there is no solution to our optimization problem. When end node memory coherence time is above $3.5$ $m$s, as we increase the coherence time of memories at repeaters, we can handle longer links which could be favored by the solver over shorter links since such paths have fewer links leading to larger end-to-end fidelity. 





\subsubsection{Utility vs. Memory capacity}

\begin{figure*}
\centering
\includegraphics[scale=0.6]{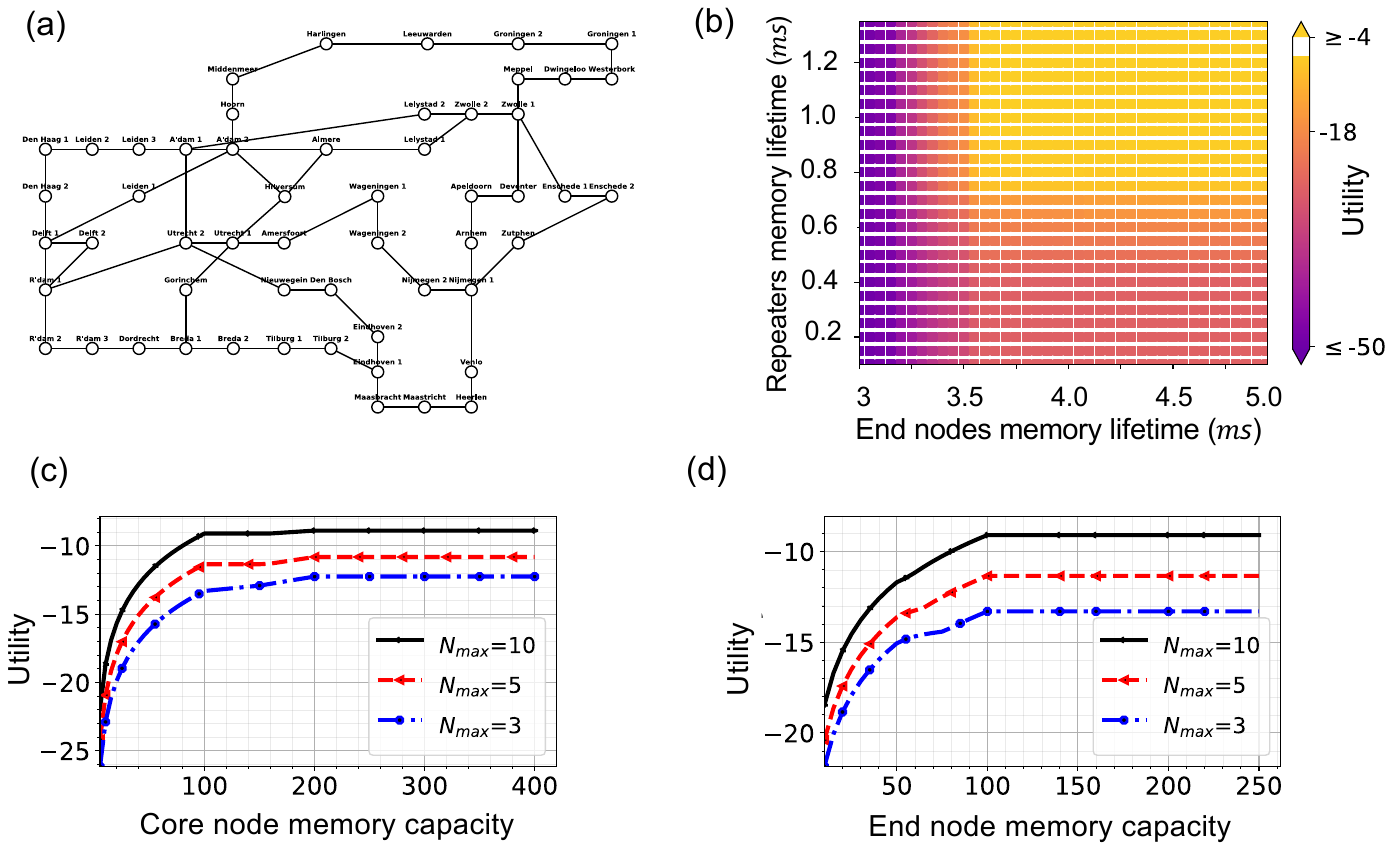}
\vspace{-0.1in}
\caption{(a) SURFnet topology, (b) utility as a function of memory coherence time for 100 different sets of $|Q|=4$ 
randomly selected user pairs, (c) utility as a function of core node memory capacity with user memory fixed {$W_E=100$}, and (d) utility as a function of end node memory capacity with fixed repeater capacity {$D=100$} for $|Q|=4$ user pairs in SURFnet.
}
\label{fig:SURFnet}
\end{figure*}

Here, we show how increasing the memory capacities of end nodes or repeaters in the network affects the maximum aggregate utility of user pairs. Figure \ref{fig:SURFnet}(c) shows that as we increase the memory capacity of the repeaters in the network (core nodes), utility increases. However, it will not affect the utility after we reach $100$ memories per end node in Figure \ref{fig:SURFnet}(c)). The same observation is true for the case the {repeater} node memory capacity is fixed at {$D=100$} and we increase the memory size at the end nodes (Figure \ref{fig:SURFnet}(d)). 

\subsubsection{Planning assumptions}

In this part, we conduct an experiment to show how different assumptions at the network planning stage can affect the performance of the network at runtime (e.g., operation time). We first plan the network for a specific workload. We call this workload the planning workload and use $Q_P$ to show the set of user pairs in the planning workload. Then, at runtime, we assume a time-slotted model where during each time slot we can have requests from a different subset of planning workload user pairs for entanglement generation. We call these workloads runtime workloads. The probability of having a user pair in each workload of runtime depends on our planning assumptions (models) that we will explore here. In this experiment, we set $D_u=100$, $E_W=200$, $q_s=1.0$, and link fidelity to $0.95$. We have $10$ user pairs in the planning workload ($|Q_P|=10$)

\textcolor{black}{We consider two different models for the runtime workloads. In the first model, we assume at each time slot, the probability of having a request for entanglement generation from each user pair is equal. We call this model the \textit{Equal probability request model}. In the second model which is called \textit{Weighted probability request model}, we assume this probability is different for each user pair. For this experiment, we assume each user pair $q$ in the planning workload has a unique weight. This weight indicates the probability of having a request from that user pair at run time at each time slot. We use $q_w$ to indicate the weight of user pair $q$. In the first model, we assume all user pairs have equal weight and that is one ($q_w=1 \quad \forall{q \in Q_P}$). In the second model, we set a different weight from the range $[0.1,1)$ for each user pair.}

\begin{figure}
\centering
\includegraphics[scale=0.7]{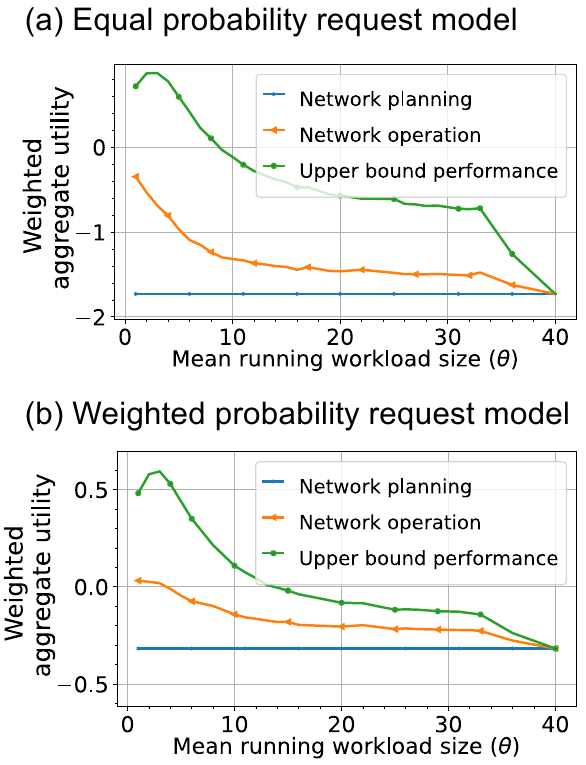}
\vspace{-0.01in}
\caption{\textcolor{black}{The impact of assumptions made during network planning on network performance during operation. When the probabilities of receiving requests at runtime from different user pairs are different, planning the network with this knowledge/assumption results in a smaller gap between the network upper bound performance and the actual performance during network operation, compared to planning without this information. }}
\label{fig:network_planning_strategies}
\end{figure}

\textcolor{black}{We first plan the network with the objective of an aggregate weighted utility function for the set of user pairs in the planning workload. The objective function is \begin{align}
\label{eq:utility}
    \sum_{q\in Q_P} q_w * \log_2 (R^q_{e2e}\cdot f(F^q_{e2e})),
\end{align}
where $F^q_{e2e}$ is the end-to-end fidelity and $R^q_{e2e}$ is the end-to-end rate for user pair $q$ respectively. We measure the performance of the network at runtime over more than $5,000$ time slots. For $5,000$ time slots, we draw $5,000$ samples from a Poisson distribution with a specific mean, $\theta$. The mean of the Poisson distribution ($\theta$) indicates the mean running time workload size. This controls the number of user pairs that we can have at each time slot. We only consider samples that are less than $|Q_P|$ and so we sample a truncated Poisson distribution. Then, for each sample that is an integer, we select those many user pairs from $Q_P$ based on the model. If $\theta$ is higher than $3.5*|Q_P|$, we choose all the user pairs for all the time slots. 
In each model and for each sample, the subset of user pairs are selected based on their weights: in the equal probability request model, pairs are chosen randomly (uniformly), while in the weighted probability request model, pairs are selected randomly but in proportion to their weights, giving pairs with higher $q_w$ values a higher chance of being chosen. Figure \ref{fig:network_planning_strategies} illustrates network performance as a function of $\theta$.
}

The green line in Fig. \ref{fig:network_planning_strategies}  indicates the case where we perform network planning for each received workload at any given time slot. We use this scheme as a reference indicating the upper bound performance, although it is unrealistic to imagine a network topology change in real-time based on the network workloads. While this approach is not practical, it shows the maximum aggregate utility that we can have for each set of user pairs at each given time if we plan the network instantaneously for the workload at that time. The blue lines correspond to the quantum network utility evaluated as the output of the optimization problem. The yellow lines show the aggregate weighted utility of the network by simulating how the demands are handled on a static network design based on the solution of the optimization problem with optimal locations for repeaters and paths for the user pairs.

Figure \ref{fig:network_planning_strategies} shows that when the probability of receiving a request at runtime from different user pairs is different, planning the network with this knowledge/assumption results in a smaller gap between the network upper bound performance and the actual performance during network operation, compared to planning without this information. As we see in Fig.~\ref{fig:network_planning_strategies} (a) and (b), the numerical values for the network planning case (blue lines) are different because we multiply the utility of each user pair with $w_q$ and in the second model $w_q<1$ for user pairs. In both cases, when we increase the mean running workload size ($\theta$) from one to $2$ and $3$, the aggregate utility increases. The reason is that with less number of user pairs some of the resources may left unused in the network as we enforce each user pair to use at most one path. Adding more user pairs can increase the aggregate weighted utility. In both models, the planned network performance is comparable with the upper bound.
For the first model, there is a higher gap between the aggregate utility at network operation time and the upper bound value compared to the second model. The reason is that in the second model, the resources have been planned to serve the user pairs with higher weight, and with a higher probability we get requests from these user pairs at run time.

\section{Related Work}
\label{sec:related}
In this section, we overview the current state of research on quantum network planning and quantum network utility maximization. \textcolor{black}{Although network planning and resource provisioning are well-established concepts in classical networks, planning for quantum networks, much like quantum routing as compared to classical routing, differs and requires distinct considerations \cite{cacciapuoti2020entanglement,shi2020concurrent,schoute2016shortcuts}.}

\myparab{Repeater placement:} Rabbie \emph{et al.}~\cite{rabbie2022designing} proposed the idea of quantum network design using the preexisting infrastructure. They formulate the problem of satisfying a certain rate and fidelity threshold for a set of user pairs with a minimum number of repeaters in the network as an optimization problem. A similar idea to our approach for network planning but with one multi-mode memory per channel has been proposed in \cite{da2023requirements}. Our work is similar to their work as we also use the preexisting infrastructure for network planning. However, our goal is to maximize the network utility which favors shortest paths with fewer repeaters.
In addition, we consider a different type of quantum memory scheme using spatial multiplexing and analyze the effect of finite coherence time of quantum memories. In contrast to Ref.~\cite{rabbie2022designing} which uses equally-distanced repeaters to estimate the end-to-end entanglement rate of a given path (regardless of the repeater positions), we evaluate the entanglement rate for each path specifically based on the exact location of the repeaters.
We further study the performance of our planned network at runtime.



\myparab{Post-planning routing:}\textcolor{black}{A branch of work in quantum networks focuses on routing and resource allocation for congestion control and entanglement generation after repeaters are placed in an area. For example, resource allocation for congestion control in quantum networks has been addressed in \cite{yang2024asynchronous,xiao2023connectionless,chen2023q}. Pant et.~al~\cite{pant2019routing} design protocols to distribute high-rate entanglement simultaneously between multiple pairs of
users. Other papers formulate routing as an optimization problem or use heuristic solutions to find optimal routes \cite{zeng2023entanglement,gu2023fendi,dai2020optimal,caleffi2017optimal,schoute2016shortcuts,li2021effective,nguyen2022multiple}. Van Meter et. al \cite{van2013path} propose new metrics for links in the network and adapt the Dijkstra algorithm for path selection in heterogeneous links networks. Chakraborty et. al \cite{chakraborty2019distributed} consider two models for the operation of a quantum network called the continuous model where link-level entanglements are continuously being generated in the background and the on-demand model where entanglement production does not commence before a request
is made. Most proposed routing protocols can be used at network operation time after the network is planned based on our proposed network planning framework. On the other hand, we can use some of the metrics proposed in the literature such as \cite{victora2020purification,caleffi2017end,van2013path} to select paths between potential places for repeaters in the planning stage. }

\myparab{Quantum network utility:} The idea of quantum network utility maximization has been introduced in \cite{vardoyan2022quantum}. Vardoyan~\emph{et al.}~\cite{vardoyan2022quantum} solve an optimization problem for finding the rates and link fidelities in order to maximize the utility function of a set of user pairs in a quantum network. They assume a centralized solver knows the topology and the location of each repeater as well as the utility function of user pairs. Their result elucidates a trade-off between the end-to-end entanglement generation rate and the fidelity.
Lee~\emph{et al.}~\cite{lee2022quantum} introduce a framework to quantify the performance and capture quantum networks' social and economic value. 
They develop an example of an aggregate utility metric for distributed quantum computing that extends the quantum volume from single quantum processors to a network of quantum processors.
Although we use a similar formula for the network utility, we are addressing a separate issue (that is the network planning). Furthermore, our approach of modeling the quantum network is quite different.
Both references model the entanglement distribution network in terms of entanglement flows along the network elementary links, which can be justified in the regime where there are infinite number of memories per channel and/or memories have infinite coherence time. In contrast, we use a physical model based on spatial multiplexing of quantum memories where the link-level entanglement generation rate can be derived explicitly based on the number of memories and the channel transmission rate. This approach in turn lets us simulate the network dynamics in an actual real-time scenario.

\section{Conclusions}
\label{sec:concl}

\textcolor{black}{This paper introduces a network planning problem as an optimization problem designed to efficiently locate quantum hardware within the existing infrastructure, aiming to maximize the utility of the quantum network. We apply our framework to several synthetic and real-world network topologies. In doing so, we illustrate the impact of memory coherence time at the repeaters and end nodes on network planning. Additionally, we analyze the influence of different fairness assumptions made during the network planning stage on network performance during runtime. We check that the qualitative trend of optimal solutions upon changing hardware parameters is as expected. For instance, we show that the coherence time requirement for quantum memories is significantly less restrictive for repeater memories compared to those of end users.}

Our optimization results on real-world examples suggest that spatial multiplexing would lead to reasonably a high end-to-end ebit rate while not imposing a huge demand on quantum memory coherence time (e.g.~sub $10$ms). A promising technology to this end is on-chip quantum memory candidates such as vacancy color centers~\cite{wang2023field}.

In the context of optimization problems, there are several avenues for future research. 
We consider a quantum network utility function based on entanglement negativity as the objective function in our optimization problem. It would be interesting to consider other objective functions for different purposes such as distributed quantum computing~\cite{lee2022quantum} or quantum key distribution~\cite{vardoyan2022quantum} and see how the optimal solution depends on the choice of the objective function.
The objective function in terms of quantum network utility is a nonlinear function in general, and to make it a linear programming we had to either drop terms or treat some variables as indices which introduces extra overhead (i.e., increases the number of decision variables). Thus, along the lines of efficiently solving the network planning problem while keeping all terms in the objective function, exploring nonlinear solvers, or reformulating the problem as a semidefinite programming could be worth pursuing. 
We should however note that either integer linear-programming or nonlinear-programming are NP-hard and our framework is only applicable to quantum networks up to a certain size.

There are also new directions to explore in network modeling and protocols.
We used an asynchronous sequential scheme for entanglement distribution. A possible direction would be to formulate network planning for other swap protocols (synchronous or asynchronous) and compare the optimal solutions across them in terms of the overall network throughput and required resources. On another note, we used a simplified model for quantum memory decoherence in terms of a hard cutoff. It would be interesting to incorporate other decoherence models also in our problem formulations.

\section{Acknowledgements}
The authors acknowledge insightful discussions with Bing Qi, Stephen DiAdamo, Matthew Turlington and Lee Sattler. This work was supported by the National Science Foundation under Grant CNS-1955744, NSF- ERC Center for Quantum Networks under Grant EEC-1941583.

{\small{
\bibliographystyle{unsrt}
\bibliography{refs} 
}}
\section{Appendix}
\subsection{Derivation of average end-to-end entanglement generation rate}
\label{app:qcast}

In this appendix, we derive the end-to-end ebit rate for a path with spatial multiplexing and explain our approximate formula.

\begin{figure}
\centering
\includegraphics[scale=0.54]{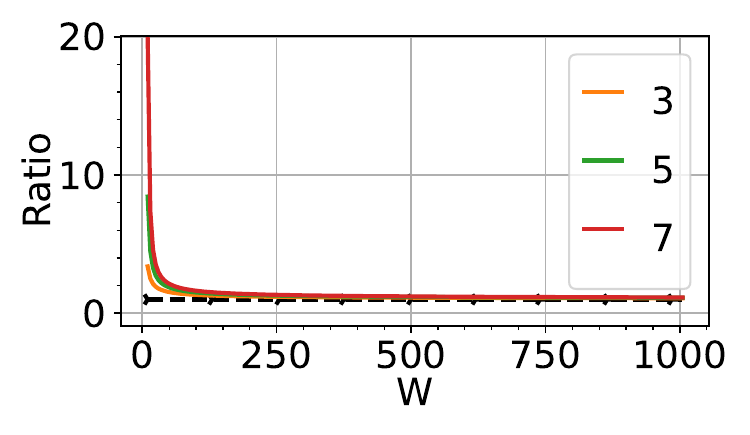}
\vspace{-0.2in}
\caption{The ratio of approximated end-to-end rate (Eq.~(\ref{eq:e2e-rate})) to exact end-to-end rate for a repeater chain. The number of repeaters on the chain is indicated in the legend. For reference, we show unit ratio as a black dashed line. The success probability of generating link-level entanglement is set at $p_i = 0.1$. 
}
\label{fig:QCAST_rate_approximation}
\end{figure}

 Consider a path with $h$ links and width $W$ where the success probability for link-level entanglement generation is  given by $p_k$ with $1 \leq k \leq h$. Let $Q^i_k$ be the probability of the $k$-th link on the path having $w$ successful ebits given by the binomial distribution $B(W,p)$ as in
 \begin{align}
    \label{eq:binomial}
    \text{Prob}(i_k = w) ={W \choose w} p^w (1 - p_k )^{W-w},
 \end{align}
where $0 \leq w \leq W$. Let $P^i_k$ be the probability of each of the first $k$ links of the path having at least $i$ successful ebits, which obeys a recurrence relation as follows
\begin{align}
\label{eq:recursive_equation2}
    P^i_k= P^i_{k-1}\cdot \text{Prob}(i_k \geq i) + \text{Prob}(i_k = i) \cdot \sum_{l=i+1}^{W}P^l_{k-1},
\end{align}
where $\text{Prob}(i_k \geq w)=1-\Phi_k(w)$ and $\Phi_k(w)$ is the CDF of the probability distribution of the $k$-th link.
The initial condition is set by the first link that is $P^i_1 = \text{Prob}(i_1 = w)$.
The average throughput can be computed by  
\begin{align}
    R_{e2e} = q_{sw}^{h-1}\sum_{w=1}^{W}i\cdot P^w_h.
\end{align}
The above expression can be computed iteratively.
Alternatively, the average throughput can be written as
\begin{align}
     R_{e2e} = q_{sw}^{h-1} \sum_{w=1}^W w \sum_{k=1}^h \text{Prob}(i_k = w)
     \prod_{j=1, j\neq k}^h \text{Prob}(i_j \geq w).
\end{align}
The binomial distribution (\ref{eq:binomial})  
in the limit $W p_k \gg 1$ can be well approximated by the normal distribution ${\cal N}(Wp_k, Wp_k(1-p_k))$ which sharply peaks at $W p_k$. 
The average throughput can then be approximated by the bottleneck link (call it $\ell$-th link) with smallest peak at $W p_\text{min}$. As a result, the dominant term in the above sum corresponds to $k=\ell$ such that  $\text{Prob}(i_j \geq w)\approx 1$ and 
$\sum_{w=1}^W w \text{Prob}(i_\ell = w) = W p_\ell$. Hence, we arrive at Eq.~(\ref{eq:e2e-rate}).
\textcolor{black}{As a numerical verification of our approximation, in Figure~\ref{fig:QCAST_rate_approximation}  we illustrate how quickly the approximate formula converges to the exact result for a linear network as we vary the number of repeaters.}

\subsection{Analysis of path-based formula}
\label{app:k-paths}

In this appendix, we show our path-based formulation with a reasonable number of shortest paths is able to find the optimal solution that the full optimization problem (the link-based formulation) can find for different random topologies with different numbers of nodes. We choose $|Q|=6$ user pairs randomly in each topology.

\begin{figure}
\centering
\includegraphics[scale=0.6]{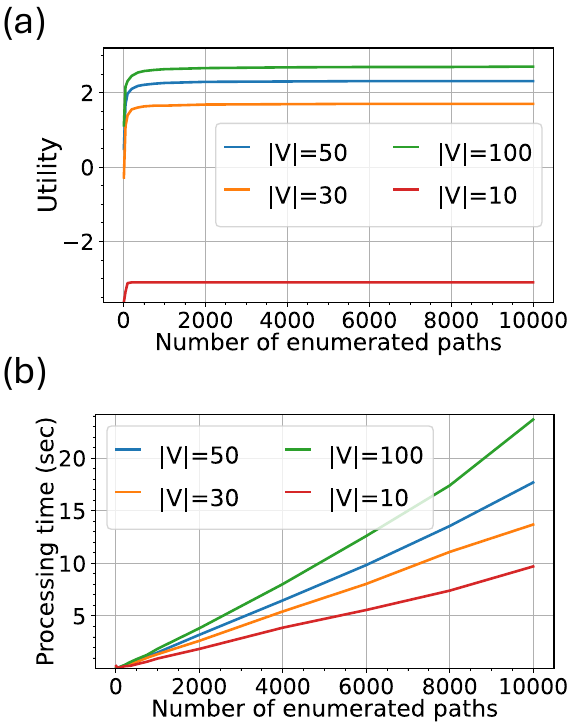}
\vspace{-0.1in}
\caption{(a) Utility and (b) the processing time as a function of the number of enumerated paths. $|V|$ is the number of nodes in random topologies.}
\label{fig:Q_utility_as_number_of_enumerated_paths}
\end{figure}

Figure \ref{fig:Q_utility_as_number_of_enumerated_paths}(a) shows that the aggregate utility of the user pairs reaches the optimal value above a certain number of enumerated paths. This is expected because if we enumerate paths from shortest to longest, the paths after a certain point will be so enough that the end-to-end rate using them drops significantly. In addition, longer paths would most likely have a larger number of links and that affects the end-to-end fidelity and the expected throughput (due to swaps). For these reasons, we set the number of enumerated paths as an input to our optimization problem in all our experiments to $4000$. Note that we also consider different versions of a path each with a different width. 

Figure \ref{fig:Q_utility_as_number_of_enumerated_paths}(b) shows the processing time in seconds for solving the path-based formulation as we increase the number of paths in the input of the optimization problem. For topologies in the size of SURFnet (with 50 nodes), we have processing time of 25 seconds when we enumerate 10k paths. We have shown in \ref{fig:Q_utility_as_number_of_enumerated_paths}(a) that enumerating only 2000 paths is enough for topologies with 50 nodes.  



\subsection{ESnet paths}
\label{app:esnet}

\begin{figure*}
    \centering
    \includegraphics[scale=0.75]{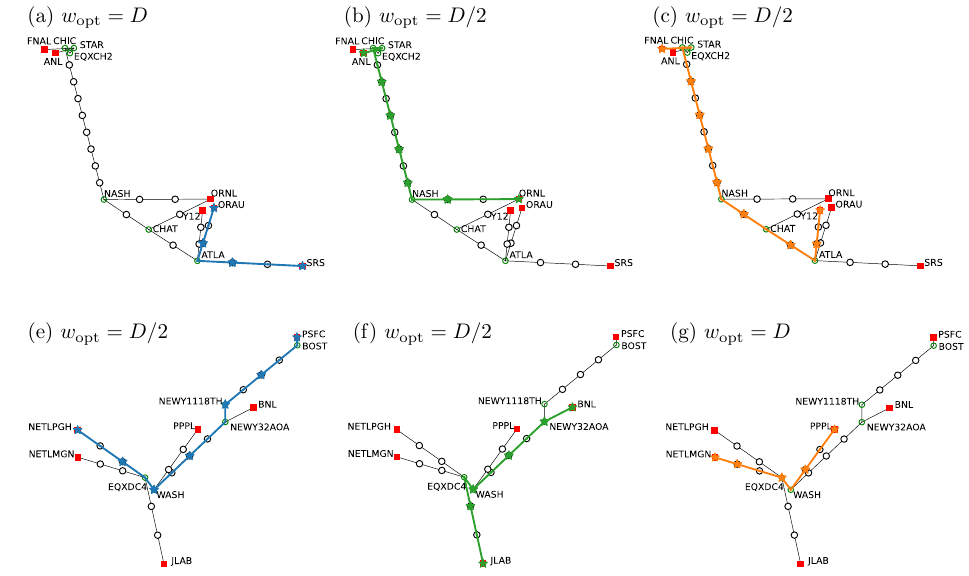}
    \caption{\protect\label{fig:esnet-paths} Optimal paths for various user pairs on  the augmented ESnet with $L_\text{max}=70$km. $w_\text{opt}$ denotes the number memories obtained for each user pair as a fraction of the repeater capacity $D$.}
\end{figure*}

Figure~\ref{fig:esnet-paths} shows the optimal paths for the user pairs on the ESnet along with the number of memories for each path in terms of repeater capacity $D$.

\begin{IEEEbiography}[{\includegraphics[width=1.1in,height=1.9in,clip,keepaspectratio]{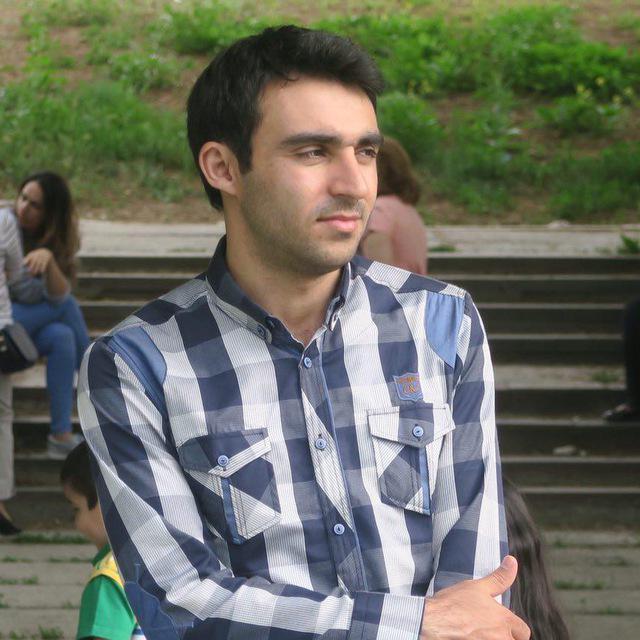}}]{Shahrooz Pouryousef} received the M.S.
degree in computer engineering from the Sharif
University of Technology, Tehran, Iran, in
2015, and the M.S. degree from the College of
Information and Computer Sciences, University
of Massachusetts Amherst, Amherst, MA, USA,
in 2021, where he is currently working toward
Ph.D. degree in computer science with the
College of Information and Computer Sciences.
His research interests include quantum
networks and distributed quantum computing.

\end{IEEEbiography}

\begin{IEEEbiography}[{\includegraphics[width=1in,height=1.25in,clip]{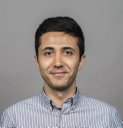}}]{Hassan Shapourian} (Member, IEEE) received
the M.S. degree in electrical engineering from Princeton University, Princeton, NJ, USA, in
2013, and the Ph.D. degree in theoretical physics
from the University of Chicago, Chicago, IL,
USA, in 2019.

He is currently a Senior Quantum Researcher
with Cisco, San Jose, CA, USA, where he leads
a wide range of projects on photonic quantum
information processing and hardware physics.

Dr. Shapourian was the recipient of several
awards including Microsoft Research Postdoctoral Fellowship, Simons
Postdoctoral Fellowship (Collaboration on Ultra-Quantum Matter), and
Kavli Institute for Theoretical Physics (KITP) Graduate Fellowship.

\end{IEEEbiography}
\begin{IEEEbiography}[{\includegraphics[width=1in,height=1.25in,clip]{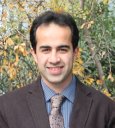}}]{Alireza Shabani} received the Ph.D. degree
in electrical engineering from the University of
Southern California, Los Angeles, CA, USA, in
2009.

He was a Postdoctoral Scholar with Princeton
University and UC-Berkeley. He founded Qulab,
a pharmaceutical startup leveraging AI to automate drug design. He was also a Senior Scientist
with Google Quantum AI Lab. He is currently
a Scientist and Entrepreneur who established a
quantum lab for Cisco Systems, San Jose, CA.
His research interests include the intersection of quantum physics, engineering, and biology.

\end{IEEEbiography}
\begin{IEEEbiography}[{\includegraphics[width=1in,height=1.25in,clip,keepaspectratio]{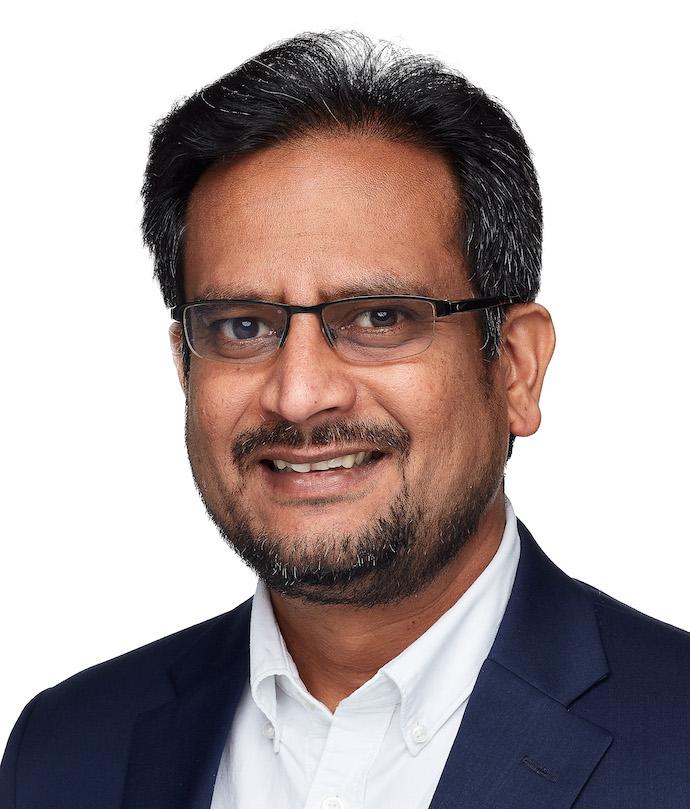}}]{Ramana Kompella}  received the M.S. degree in
computer sience and engineering from Stanford
University, Stanford, CA, USA, 2001, and the
Ph.D. degree in computer sience and engineering
from the University of California San Diego, San
Diego, CA, in 2007.

His background is a perfect confluence of research and entrepreneurship, having spent a significant part of his career in academia as a tenured
faculty with Purdue University, as well as in startups as a Cofounder and Chief Technology Officer (CTO) with AppFormix and a Cofounder, Head of Engineering, and
CTO with Candid alpha project inside Cisco. He was also part of the
Google’s network architecture team where he focused on large-scale data
center network operations. He is currently the Head with Cisco Research,
San Jose, CA.

\end{IEEEbiography}

\begin{IEEEbiography}[{\includegraphics[width=1in,height=1.25in,clip,keepaspectratio]{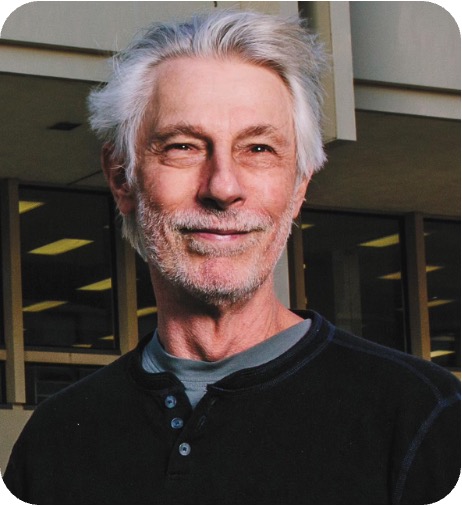}}]{Don Towsley} (Life Fellow, IEEE) received
the Ph.D. degree in computer science from the
University of Texas, Austin, TX, USA, in 1975.

He is currently a Distinguished Professor with
the Manning College of Information and Computer Sciences, Amherst, MA, USA. His research
interests include performance modeling and analysis, and quantum networking.

Dr. Towsley was the recipient of received several achievement awards including the 2007 IEEE
Koji Kobayashi Award and the 2011 INFOCOM
Achievement Award.

\end{IEEEbiography}

\EOD

\end{document}